# Intelligent Internal Temperature Control of Food in Standard Convection Ovens

**Ramon Gloumakov,** *Student Member, University of Michigan – Dearborn*

**ABSTRACT**

This paper introduces a feedback-based temperature controller design for intelligent regulation of food internal temperature inside of standard convection ovens. Typical convection ovens employ an open-loop control system that requires a person to estimate the amount of time needed to cook foods in order to achieve desired internal temperatures. This approach, however, can result in undesired results with the final food temperatures being too high or too low due to the inherent difficulty in accurately predicting required cooking times without continuously measuring internal states and accounting for noise in the system. By implementing and introducing a feedback controller with a full-order Luenberger observer to create a closed-loop system, an oven can be instrumented to measure and regulate the internal temperatures of food in order to automatically control oven heat output and confidently achieve desired results.

**INTRODUCTION**

Ovens are one of the several commonplace cooking appliances, along with refrigerators and stoves, that are practically in all U.S. households with approximately 99% of said households having them [1]. Convection ovens specifically are being increasingly promoted over conventional ovens since they employ an improved air circulation system that allows for more even high temperature cooking. The goal is to enable users to confidently cook foods to reach predefined temperatures and conditions. Frequently, safe and recommended temperature guidelines can be found in various forums and recipes as well as in guidelines coming from government agencies such as the CDC and FDA. See a list of example foods and their corresponding recommended final temperatures in Table 1, some of which were used in the control, analysis, and design [2][3].

| Food | Safe Temperatures (°F) | Recommended Temperatures (°F) |
|---|---|---|
| Steak | 145 | 130 – 135 |
| Chicken | 165 | 165 – 175 |
| Turkey | 165 | 165 – 175 |
| Seafood | 145 | 130 – 140 |
| Bread | 140 | 180 – 200 |
| Potato | 140 | 200 |

*Table 1: Internal Temperatures Guidelines for Various Commonplace Foods*

Although the final desired temperatures might be known, an average convection oven user will find it difficult to consistently achieve these temperatures. Without being able to reliably estimate the precise required cooking time or being able to reliably adjust oven output to account for inherent "noise" in the system, the final results will be more susceptible to error. This cooking method is consequently, in-part, an open-loop system, one that does not monitor the internal states of the food it is cooking and does not take them into account as part of a feedback system. To improve this technology, a novel feedback controller can be introduced in order to create a closed-loop system, one that actively monitors the food temperature and can adjust the oven heat output to ultimately achieve a predefined output. In addition, the feedback controller can be supplemented with a Luenberger observer that indirectly estimates the state temperatures of the system to be used by the feedback controller. This avoids having to introduce additional sensors and cost to the design. The next section will discuss the models and equations used for designing the state-space representation of the proposed controller.

## FUNDAMENTALS OF HEAT TRANSFER

Heat transfer can be broken down into three different types of mechanisms: conduction, convection, and radiation. Conduction is the transfer of heat when particles are in direct contact with each other. Convection is the transfer of heat through a fluid or medium, such as water or air. Radiation is the transfer of heat through electromagnetic waves that are emitted by hotter objects and absorbed by cooler objects. The basic formulas for calculating the amount of heat transfer occurring in a system for each of these mechanisms can be found in *Perry's Chemical Engineers' Handbook* [4]. For the purposes of this study, it is assumed that the primary source of heat transfer in a convection oven is convection, with the other two assumed negligible. For convective heat transfer, the primary equation is shown in Equation (1) [4].

$$Q = mC_p \frac{dT}{dt} = h_c A(T_i - T_j) \quad (1)$$

Q: Amount of Heat Transferred (Btu)
m: Mass of Object (lb)
$C_p$: Specific Heat Capacity (Btu/(lb °F))
$h_c$: Convective Heat Transfer Coefficient (Btu/(ft² hr °F))
A: Heat Transfer Surface Area (ft²)
$T_i$: Temperature of Object *i* (°F)
$T_j$: Temperature of Object *j* (°F)

Most of these values can be measured real time or looked up for various items, but the heat transfer coefficient ($h_c$) values depend on the surface characteristics of the objects or fluids that are in contact. These values, unfortunately, are not readily available for all materials that might be in contact with air, including the various foods and materials that take part in an oven's cooking process. Instead, these values can be estimated by using empirical formulas. For this study, by assuming that all heat transfer is occurring between air and another isothermal flat surface, heat transfer coefficients were estimated using Equations (2)-(6) [4].

$$Gr = \frac{D^3 \rho^2 g \Delta T \beta}{\mu^2} \quad (2)$$

$$Pr = \frac{\mu C_p}{k} \quad (3)$$

$$Nu = 0.138 * Gr^{0.36}(Pr^{0.175} - 0.55)^{0.25} \;;$$
for Gr > $10^9$ \quad (4)

$$Nu = 0.683 * Gr^{0.25} Pr^{0.25} \left(\frac{Pr}{0.861 + Pr}\right)^{0.25} \;;$$
For Gr < $10^9$ \quad (5)

$$h_c = \frac{Nu \, k}{D} \quad (6)$$

Gr: Grashof Number (dimensionless)
Pr: Prandtl Number (dimensionless)
Nu: Nusselt Number (dimensionless)
D: Characteristic Surface Length (ft)
k: Fluid Thermal Conductivity (Btu/ft s °F)
ρ: Fluid Density (lb/ft³)
g: Gravity (ft/s²)
β: Fluid Expansion Coefficient (1/°F)
μ: Fluid Viscosity (lb_f s/ft²)

By using these equations and the fluid properties of air at an assumed temperature point, approximate values for the heat transfer coefficients between the air and oven walls as well as the air and typical foods can be calculated. With these formulas and values, a state-space representation was generated for the system as summarized in the next section.

## SYSTEM STATE-SPACE REPRESENTATION

To generate a state-space representation of the oven system, it is important to distinguish what is taking part in the heat transfer process. There are three materials in play: the food, the oven walls, and the air inside. Therefore, there are three system states taking part in the design. These states are affected by the heat transfer occurring between each of the surfaces as well as the heat input produced by the oven itself. This incoming heat is assumed to go directly and exclusively into the air circulating the oven and is calculated based on the oven initial preheated temperature. The output of the system is the measured internal temperature of the food that will ultimately be used for feedback in the controller. By applying and rearranging Equation (1) from the previous section, the following state-space Equations (7)-(10) can be generated.

$$\dot{T}_{air} = \frac{1}{m_{air}C_{p,a}}(m_{air}C_{p,a}(T_i - T_{air}) - h_{a,w}A_{a,w}(T_{air} - T_{wall}) - h_{a,f}A_{a,f}(T_{air} - T_{food})) \quad (7)$$

$$\dot{T}_{wall} = \frac{1}{m_{wall}C_{p,w}}\left(h_{a,w}A_{a,w}(T_{air} - T_{wall})\right) \quad (8)$$

$$\dot{T}_{food} = \frac{1}{m_{food}C_{p,f}}\left(h_{a,f}A_{a,f}(T_{air} - T_{food})\right) \quad (9)$$

$$y = T_{food} \quad (10)$$

$T_i$: Oven Initial Preheated Temperature (°F)
$h_{a,x}$: Air, Object 'X' Heat Transfer Coefficient
$A_{a,x}$: Air, Object 'X' Surface Area (ft$^2$)
$\dot{T}$: Derivative of Temperature (dT/dt)
$y$: State-Space Output (°F)

By using the material properties of the three components of the system and assuming 26in x 16in x 16in dimensions for a standard household oven, equations above can be further simplified and a state-space model can be generated with the moving variables. These properties assuming an initial state temperature of 80°F can be found in Table 2 and 3 [5][6]. Since material properties will vary between foods, this study only covers oven control for steak, chicken, and potatoes.

|  | Air |
|---|---|
| $C_{p,a}$ | 7.731 |
| $\rho$ | 2.284x10$^{-3}$ |
| $k$ | 4.233x10$^{-6}$ |
| $\beta$ | 1.87x10$^{-3}$ |
| $\mu$ | 3.852x10$^{-7}$ |

*Table 2: List of Air Properties*

|  | Wall | Steak | Chicken | Potato |
|---|---|---|---|---|
| $C_{p,x}$ | 0.22 | 0.66 | 0.77 | 0.82 |
| D | 2.0 | 0.5 | 0.5 | 0.3 |
| $h_{a,x}$ | 1.069 | 1.189 | 1.189 | 1.141 |

*Table 3: Other Material and Surface Properties*

With all the material properties accounted for and the heat transfer coefficients calculated (last line in Table 3) using the properties and Equations (2)-(6), all that is left is to calculate the mass and surface area of each material. These values could in theory be measured using load cells and displacement sensors, but for this project the values were estimated using material densities and typical material dimensions (see Table 4) [7][8].

|  | Air | Wall | Chicken | Steak | Potato |
|---|---|---|---|---|---|
| m | 0.283 | 75.0 | 0.5 | 0.5 | 0.375 |
| $A_{a,x}$ | -- | 15.11 | 0.375 | 0.375 | 0.256 |

*Table 4: Mass and Surface Area Values*

Using Tables 2 through 4, values can be entered into Equations (7)-(10) and a final state-space representation of the system can be generated for each of the three food choices. Equations (11)-(13) will be the primary set of equations used in the controller design, simulation, and analysis sections that will be discussed next.

$$\begin{bmatrix} \dot{T}_{air} \\ \dot{T}_{wall} \\ \dot{T}_{steak} \end{bmatrix} = \begin{bmatrix} -8.587 & 7.383 & 0.204 \\ 0.979 & -0.979 & 0 \\ 1.351 & 0 & -1.351 \end{bmatrix} \begin{bmatrix} T_{air} \\ T_{wall} \\ T_{steak} \end{bmatrix} + \begin{bmatrix} 1 \\ 0 \\ 0 \end{bmatrix} Q_i$$

$$y_{steak} = \begin{bmatrix} 0 & 0 & 1 \end{bmatrix} \begin{bmatrix} T_{air} \\ T_{wall} \\ T_{steak} \end{bmatrix} \quad (11)$$

$$\begin{bmatrix} \dot{T}_{air} \\ \dot{T}_{wall} \\ \dot{T}_{chicken} \end{bmatrix} = \begin{bmatrix} -8.587 & 7.383 & 0.204 \\ 0.979 & -0.979 & 0 \\ 1.158 & 0 & -1.158 \end{bmatrix} \begin{bmatrix} T_{air} \\ T_{wall} \\ T_{chickn} \end{bmatrix} + \begin{bmatrix} 1 \\ 0 \\ 0 \end{bmatrix} Q_i$$

$$y_{chicken} = \begin{bmatrix} 0 & 0 & 1 \end{bmatrix} \begin{bmatrix} T_{air} \\ T_{wall} \\ T_{chicken} \end{bmatrix} \quad (12)$$

$$\begin{bmatrix} \dot{T}_{air} \\ \dot{T}_{wall} \\ \dot{T}_{potato} \end{bmatrix} = \begin{bmatrix} -8.516 & 7.383 & 0.134 \\ 0.979 & -0.979 & 0 \\ 0.950 & 0 & -0.950 \end{bmatrix} \begin{bmatrix} T_{air} \\ T_{wall} \\ T_{potato} \end{bmatrix} + \begin{bmatrix} 1 \\ 0 \\ 0 \end{bmatrix} Q_i$$

$$y_{potato} = \begin{bmatrix} 0 & 0 & 1 \end{bmatrix} \begin{bmatrix} T_{air} \\ T_{wall} \\ T_{potato} \end{bmatrix} \quad (13)$$

## OPEN-LOOP SYSTEM BEHAVIOR

Using the state-space models identified in the previous section, open-loop behavior can be analyzed. First, system stability can be evaluated by calculating the system poles of each set of equations. This can be done using the Matlab command eig() [9]. The eigenvalues or poles of each set is shown in Table 5.

| Eigenvalue | Steak | Chicken | Potato |
|---|---|---|---|
| Pole 1 | -9.472 | -9.467 | -9.390 |
| Pole 2 | -0.104 | -0.104 | -0.104 |
| Pole 3 | -1.341 | -1.153 | -0.951 |

*Table 5: Oven System Eigenvalues / Poles*

The first detail to notice in Table 5 is that all poles are negative, resulting in asymptotically stable systems. This makes sense since the open-loop system temperatures of the oven and food will simply converge overtime to the user-defined preheated temperature. Naturally, the system temperatures wouldn't suddenly become unstable on their own if the oven is held at one heat input. Although the system is stable, if left unregulated, the temperature of the food will increase to match the oven temperature and will exceed the desired values listed in Table 1. By using Matlab command lsim() and Equations (11)-(13), this open-loop behavior can be visualized for the three food items, assuming an oven preheated temperature of 400°F and food initial temperature of 80°F. Figure 1 shows the open-loop behavior for the steak model with the other two foods exhibiting very similar results.

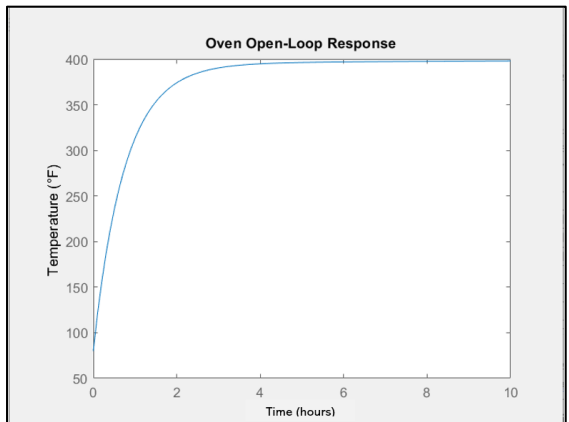

*Figure 1: Oven and Steak Open-Loop System*

## CLOSED-LOOP CONTROLLER DESIGN

In order to control oven heat input and ensure the internal temperatures of food never exceed values listed in Table 3, feedback controllers can be introduced to the systems. To design these controllers, this study employs pole placement via state feedback method with a full-order observer to compute the final controller gain matrix and observer output injection matrix [10][11]. By varying the pole placements, system dynamics and behaviors can be evaluated and tuned until desired performances are realized. For the purposes of this study, observer poles were at values at least 5 times more negative than the controller poles in order to ensure accurate estimation of system states for feedback. Through the process of iteration and the use of Matlab command place(), the system poles in Table 6 were selected to ensure food temperatures quickly reach the desired values in Table 1 without exceeding them [9].

|  | Steak | Chicken | Potato |
|---|---|---|---|
| Final Temp (°F) | 135 | 165 | 200 |
| Controller Pole 1 | -39.0 | -27.0 | -38.5 |
| Controller Pole 2 | -0.1 | -0.1 | -0.05 |
| Controller Pole 3 | -1.0 | -1.0 | -0.97 |
| Observer Pole 1 | -195.0 | -135.0 | -192.5 |
| Observer Pole 2 | -0.5 | -0.5 | -0.25 |
| Observer Pole 3 | -5.0 | -5.0 | -4.85 |

*Table 6: Controller and Observer Pole Design*

With the desired system poles identified, the controller gain (K) and observer output injection (L) matrices are calculated using Matlab place() command. These matrices can then be re-introduced to the original state-space equations to reach the final system equations with added controllers and observers. The final equations were calculated using Equations (14)-(16).

$$A_{feedback} = \begin{bmatrix} A - B*K & B*K \\ zeros(size(A)) & A - L*C \end{bmatrix} \quad (14)$$

$$B_{feedback} = \begin{bmatrix} B \\ zeros(size(B)) \end{bmatrix} \quad (15)$$

$$C_{feedback} = [C \quad zeros(size(C)] \quad (16)$$

With the final equations generated for the three food items, the closed-loop system behaviors can be visualized using Matlab lsim() command. Figures 2 through 4 confirm the final temperatures converge to and never exceed the recommended values listed in Table 1 using the poles identified in Table 6. In addition, the figures show other temperature response times if the pole locations were varied, reduced, or magnified by various amounts. The figures illustrate performances with both faster and slower response times as well as responses that exhibit overshoot and undershoot, both of which are not desired in this proposed oven operation.

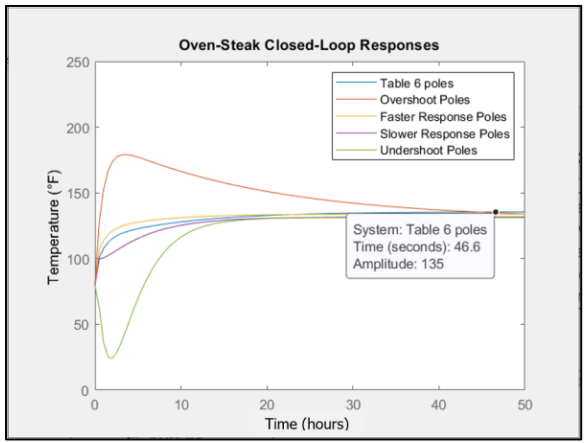

*Figure 2: Oven and Steak Closed-Loop System*

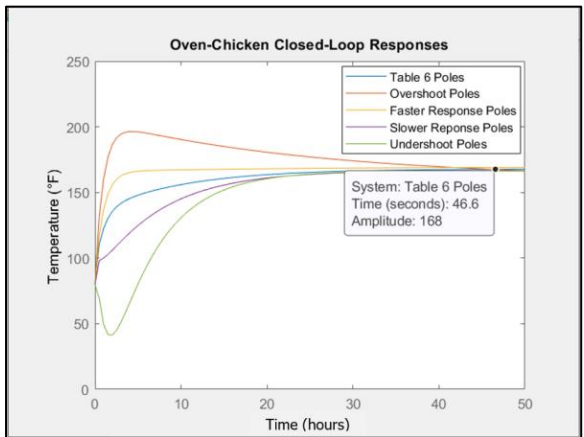

*Figure 3: Oven and Chicken Closed-Loop System*

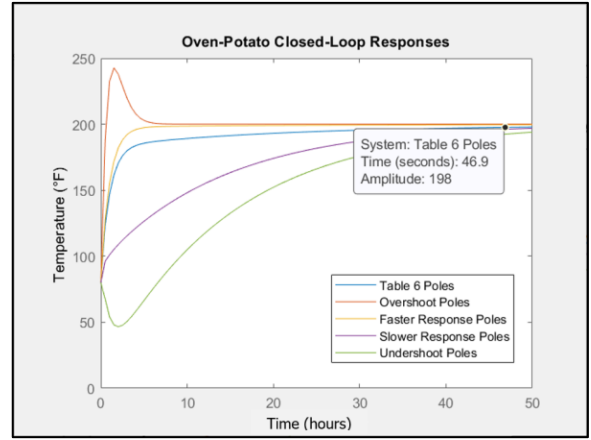

*Figure 4: Oven and Potato Closed-Loop System*

## CONCLUSION

To improve internal temperature control of food inside standard convection ovens, a feedback controller with a full-state observer can be utilized. By designing and integrating an oven controller that continuously measures internal temperature, the oven heat output can be actively controlled in order to achieve final desired food temperatures without outside intervention.

In this paper, controller examples were designed for three foods: steak, chicken, and potatoes. By employing the pole placement via state feedback method to select desired poles, the final results were simulated and tuned to achieve desired performances and outputs. Each design ensured that the final simulated temperatures converged to and never exceeded the temperature values recommended by various food forums and government agencies. To design controllers for a wider variety of food, additional logic would likely need to be investigated and added with this paper providing the necessary basic principles for further implementation. The proposed controller design can also be further enhanced by exploring component cost, controller efficiency, oven instrumentation, and potential unexpected noise to ensure optimal controller behavior. As future extension for this design, a Linear-Quadratic-Regulator (LQR) or Linear-Quadratic-Gaussian (LQG) controllers can be investigated.